\documentclass[sigconf,10pt,nonacm]{acmart}
\usepackage{url}

\title{What if we could hot swap our Biometrics?}

\author{Jon Crowcroft}
\affiliation{%
  \institution{University of Cambridge}%
  \city{Cambridge, England}
  \country{UK}%
}

\author{Anil Madhavapeddy}
\affiliation{%
  \institution{University of Cambridge}%
  \city{Cambridge, England}
  \country{UK}%
}

\author{Richard Mortier}
\affiliation{%
  \institution{University of Cambridge}%
  \city{Cambridge, England}
  \country{UK}%
}

\author{Chris Hicks}
\affiliation{%
  \institution{Turing Institute}%
  \city{London, England}
  \country{UK}%
}

\author{Vasilios Mavroudis}
\affiliation{%
  \institution{Turing Institute}%
  \city{London, England}
  \country{UK}%
}

\begin{abstract}
What if you could really revoke your actual biometric identity, and install a new one, by live rewriting your biological self?

We propose some novel mechanisms for hot swapping identity based in novel biotechnology. We discuss the potential positive use cases, and negative consequences if such technology was to become available and affordable.

Biometrics are selected on the basis that they are supposed to be unfakeable, or at least not at reasonable cost.  If they become easier to fake,  it may be
much cheaper to fake someone else's biometrics than it is for you to change
your own biometrics if someone does copy yours. This potentially makes 
biometrics a bad trade-off for the user.

At the time of writing, this threat is highly speculative, but we believe it is (a bit like post-quantum crypto) worth raising and considering the potential consequences.

\end{abstract}

\begin{document}

\maketitle

\section{Both Foundational and Functional Identity}

The Internet connects around ten billion people and systems. 
One of the big problems with this scale is we need to know who you are, and we can't just rely on you being vouched for by some nearby friends, family or colleagues, apparently~\cite{ford2020identity,sheng2006johnny}.

There are two leading approaches to issuing electronic credentials that can be used to address the problem of remote authentication: \emph{foundational} and \emph{functional} identities. Depending where you are in the world one or the other of these will be most familiar. Foundational ID systems, ``popularised'' by 1.4 billion people enrolled in India's eponymous Aadhaar~\cite{aadhaar_dashboard}, are general-purpose identities used for a wide range of activities (e.g., a passport used for travel, age verification, bank account opening, conveyancing etc) whereas functional identities are designed and built for a specific purpose (e.g., national health service number).

Unique, foundational identity is typically rooted in unique biological markers like
fingerprints, retina, iris, less so face, and even less so behaviour, and, 
of course, your DNA.

Biometrics are increasingly common in proving who a person is to a device, usually through a secure sensor (fingerprint reader, or camera in a secure mode, with a secure channel) encrypted at or near source, and then used to sign communication to authorise according to some attached credentials~\cite{applePlatformSecurity2024}. Data minimisation principles hopefully being used, things like ``age verification'' only reveal a binary fact (``this person is over 21'') rather than an actual birth date. 

The split between foundational (just unique identity) and functional (establishing metadata associated with credentials) is now fairly standard, and the use of fancy biometrics is often carefully limited to the former function, whereas the latter can be revealed from a previously authorized device, associated with the identified user, assured via cryptographic means.

Revealing raw biometric data is regarded as incredibly risky, since once it is compromised, the corresponding unique biology cannot be used again. Biometrics are close to impossible to change, but may be relatively easy to imitate (e.g. fingerprints rendered using glue, face simply by copying photos even 2.5D deep fake copies). 

What if you could modify a biometric? You could use it directly, and simply revoke it and update as needed.

In this paper we present three ideas which may seem somewhat like science fiction at this point in time, but we offer as a thought experiment towards what might be possible in the not too distant future.

\section{Idea 1 - papers please}

The requirement to carry proof of identity is relatively recent - the widespread use
of passports which carry a photograph of the holder dates from early 20th century.
Digital identity is increasingly commonplace, for example with the use of mobile driving licenses, eliminating the need for paper.

Researchers have proposed means to verify credentials in general via mobile services, for example using secure processing on the SIM on a feature phone\cite{hicksSIMpleID22}, further reducing costs and generalising the functionality.

We would like to eliminate the use of devices altogether, and store foundational and
functional identity attributes directly within the subject, and then communicate these securely.

One approach might employ mutable bar codes displayed on the skin.
Animated tattoos have appeared in SF frequently, and some technology is even the subject of a startup\cite{tattoo}.

\begin{figure}
    \includegraphics[width=0.5\linewidth]{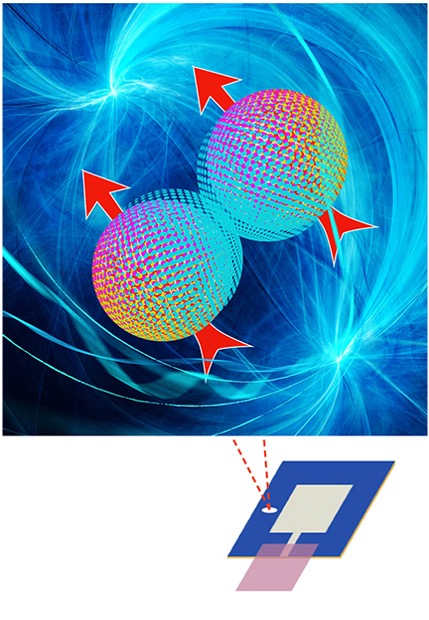}
    \caption{SQUID: Super-conducting Quantum Interference Devices}\label{fig:squid}
    \Description[<SQUID>]{<Super-Conducting Quantum Interfering Devices>} 
    \end{figure}

\begin{figure}
    \includegraphics[width=0.5\linewidth]{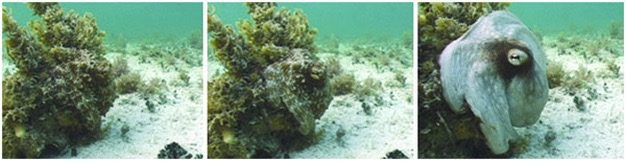}
    \caption{Chromatophores in Action}\label{fig:chromatophores}
    \Description[<Squid using Camoflage>]{<Squid using Chromatophores to camoflage itself>} 
    \end{figure}

These would be digitally signed, and only reveal what was necessary to gain access to some 
service (``relying party'') -- data minimisation would be under your control, and talk about self sovereign!
The technology could derive from squid - no, not this kind as in figure \ref{fig:squid}, although we will come back to that\cite{SQUID}, but related to this kind of squid feature, illustrated in action in figure \ref{fig:chromatophores}.

There are two seperate technical problems. Firstly, you have to splice the actual gene tech for chromatophores into humans, e.g. using CRISPR-CAS9 tech\cite{CRISPR}. Secondly, you need to interface the human autonomic nervous system to the newly acquired dynamically updatable {\it tattoos}.
These two challenges may take a while to solve, but be assured, the non-peacetime applications might drive rapid development of solutions.

We note that squid react to their environment with remarkable speed because much of the chromoatophore control is local (the squid have distributed intelligence too, which may be related). 
To make this work for our purposes, we will need to modify this mechanism to operate either under central control (brain) or replicate relevant input to local nervous system more like the squid. Enhanced, decentralised reflex-like intelligence would also be an asset in a human, for example, having applications like advanced muscle-memory, very handy for playing musical instruments or sports (including online games), especially given the relatively slow signal propagation of voluntary actions through the human nervous system when compared to localised peripheral mechanisms (e.g., thermally-induced nociceptive withdrawal reflex muscle contractions which occur much more quickly)~\cite{JURE2019259}. 

Indeed, the human nervous system comprises many intricate and varied processes that may either be repurposed for the controlling of chromatophores or towards alternative mechanisms for storing and displaying identity attributes. For example, fingertip ``pruning'' upon submersion in water is controlled by the ulnar nerve~\cite{ORiain1973}. Perhaps enhancement of ulnar nerve function could one day facilitate intentional, user-controlled variations in digital fingerprints for remote identification? Promisingly, horripilation (a.k.a goosebumps) is not limited to cold temperate exposure and can occur in response to a wide variety of emotional states~\cite{MCPHETRES20226}. Since horripilation occurs with activation of the sympathetic nervous system, also affecting skin conductance, there are many rudiments from which future scientists and bio-hackers may implement dynamic biological authentication displays.

%might augmentation of the sympathetic nervous system responsible for controlling horripilation (a.k.a goosebumps) allow for more granular control of 

%Humans do have a distributed nervous system / intelligence, and some of it works very quickly (e.g., heat reaction). Wrinkling/pruning of the
%fingers is controlled by the nervous system which is very interesting for the
%purposes of this paper. Perhaps we can repurpose these functions to controlling the chromatophores.

%Chris suggests perhaps goose bumps could be interesting since they run in a dot pattern which, if we could control it, might enable specific information or codes to be {\it displayed}.

We also need to make sure that the channel from attribute data (e.g.  verifiable credentials) to the display/output is secure, so that tampering with the signed visible data isnt feasible - this is already part of  today's biometric readers (fingerprint etc) and subject to NIST  standards including verification procedures.

Watching TV on your hands could be a much later development, but chromatophores certainly have the capability for highly dynamic rendering. On an historical note, when the Ethernet was first deployed, devices shipped with one hard-wired address. Fairly quickly this was made mutable, and more recently, MAC addresses were randomly cycled, to prevent tracking of devices over space and time. One can imagine cycling through randomly generated facial appearances to provide the same mitigation of intrusive surveillance.

So far, we've only talked about mutable {\it representations} of identity. What if we could modify the actual root of our biological self? Lets look at this next.

\section{Idea 2 - re-write your retina}

What if you could re-write your retina, your iris\cite{iris}, your fingerprint, or even your face?  Of course, people have temporarily 
overridden their fingerprints\cite{fingerprint}, or just worn a mask, but we're discussing actual replacement of the echt biological matter.

RNA/transcriptase etc (as per teaching immune system to recognise foes) is an affordable mechanism for delivering new information and functionality into a biological entity.

But now we face this problem: how do you know a person is still the same person? depends whether we go as far as re-writing all the DNA or leave the major portion of it alone.

Note RNA printing  was posited during the creation of vaccines during the recent pandemic, where mRNA \cite{mRNA} was used to instruct your
cells to create proteins like the actual virus, that would then train your immune system to respond to this intruder.
mRNA consists of a long sequence of 4 proteins (lke DNA) \url{https://rdcu.be/esSFw} which can be prepared and made available like 4 colours of ink.

We can use this to store attributes (citizenship, entitlements, exam results/qualifications,  etc) in junk DNA, which can then be read out and verified by a relying party.

So now we have embedded both foundational and functional identity within your body. What about communication? Can we now provide privacy and non-repudiation between people again employing bio-technological means? We look at that next.

\subsection{Idea 2.5 - Hybrid}
While rewriting biological identity may be far-future, we can imagine a transitional model where biological traits are bound to external components to form a cryptographic credential. Consider an individual's iris: immutable, unique, but also vulnerable to capture or cloning.

For this we would need composite lens so as to form a split credential where part of the identity comes from the unaltered biological pattern (e.g., iris structure), and part comes from the wearable.
Only when both are present can the full credential be reconstructed and authenticated. This is analogous to threshold cryptography, where no single party has all the information needed to perform sensitive computation or verification alone.

This offers revocability by replacing the device component without touching the biological base. Also stealing the lens without the correct eye is pointless while the eye without the lens reveals nothing useful to an attacker. The system can be designed to emit only proofs (zkp) that an individual meets a requirement (e.g., access level), without revealing underlying attributes. The lens can incorporate a nonce-based rotation (e.g., daily or weekly replacement) preventing replay attacks if biometric data is captured.
This dual-track identity mechanism opens up a spectrum of flexible identity management options where full biological rewriting isn't yet feasible or desirable. It also creates a more graceful path from current device-bound digital ID to future biologically embedded credentials.

\section{Idea 3 - the honest smell}

How about secure communication directly between pairs of humans (or any other beings - e.g. human and pet or livestock)?

Launching from the assumption that an individual can now create signed verifiable credentials
biologically, we add one more technical suggestion to the mix, which is to leverage ideas from Quantum Key Distribution (QKD)\cite{QKD}, and observe that at least some scientists have suggested that biological sensing can detect quantum level effects.

The idea here would be to provide secure communication without third party key distribution services. Instead, we suggest that individuals could continue to identify themselves, but also exchange keys directly using (for example) pheromones\cite{smell}.

Using QKD, we can deliver tamper proof pairwise key exchange. This
needs entanglement -- we are not sure if this is part of current biological quantum effect, and also generating pairs of entangled particles biologically might be tricky. It is also possible that it might work better with taste than smell. Certainly, at the receiving end, there is e-nose technology that might help, so it is the transmission side that is a challenge. Again, some hybrid human-technology solution might be applicable. I would propose using a protocol such as Stajano's Resurrecting Duckling\cite{10.5555/647218.720721} for the actual setup.

This is why idea 3 also needs further research.

% why don’t we use existing methods at this point? Coming up with secure keys is the hard part. Getting weird about it how about we shake hands and some bioengineered virus (long term) or bacteria (short term) is constructed from our junk DNA and used to encrypt future communication (especially facilitated by bio-implant/chips that can encrypt and synthesise our voice using the keys)

\section{Threats}

% Dare I say some suggestive attacker model(s) could be indicated? Might be worth sketching some even if they don’t make the paper just to structure the proposed ideas.

The core of ideas 1 \& 2 undermines existing identification, since one can use mutable id to impersonate someone. This undermines the use of unique id for a variety of services (voting, payment, or forensics, just for a few obvious contexts).

The use of socially constructed identity might be attractive in terms of human-to-human trust, and idea 3 is supposed to help support that. However, legitimate reasons to surveil individuals and their communication would be severely curtailed.

Many technical threats exist to the proposed techniques, not least assurance that the service does what it claims to; and how do you know the people operating such a service are who they claim to be? On the other hand, at least one group of users in the community might welcome the chance to modify their biometrics, and that is undercover spies who wish to carry out multiple operations in multiple countries, whilst masquerading under different cover identities\cite{spies}.

\section{One Possible Solution Space}

Socially constructed identity (proof of human personhood via human interaction with friends and family) could be a way to build a complex, behavioural, multi-modal biometric, which would include much interaction and therefore require entirely synthetic humans to fool. 

This has actually been used for remote onboarding and human attestation in Ethiopia in their government national digital Identity service, fayda\cite{fayda}. 

Once a synthetic being is feasible at this level of fidelity\cite{trigger}, perhaps one is no longer so worried about unique identity. Other problems may be a priority. However, if the quantum key distribution technique using smell described in the previous section was used, then the synthetic human would have different keys from the {\it rea;} human. In practice, the entangled pair stage might need additional hardware support if we cannot solve the biological pair creation stage, and those devices could, of course, be vulnerable to attack.

Graph properties have also been used in networks of devices to provide a likely unique signature, e.g. to mitigate spam generated by clones, for example in the sybilguard\cite{10.1145/1151659.1159945} system.
\section{Conclusion}

What is identity? We don't ask this as a philosophical question but as a real technical challenge. We assume certain characteristics of humans are immutable over their lifetime. This may not always be true. If real-world metrics that distinguish one individual from another are modifiable, this can have both positive and negative impacts on how we deal with assurance about social and economic rights and obligations. The pace of change in science means that these impacts may not be so far off, and as with other technologies such as AI and Quantum Computing, we should be prepared.

One alternative approach to assurance is to use socially constructed identity, where the graph of other people who vouch for an individual {\em is} their identity. Some serious uses of this  include the onboarding of people in remote villages in the Ethiopian national government id system {\it Fayda} (see \url{https://en.wikipedia.org/wiki/Fayda_ID}), which needs to deal with extreme challenges of inclusivity. A misleading variant of this in the digital domain is this other proof-of-personhood\cite{7966966}, which depends on decentralisation of technology rather than the natural social federation of actual humans.

\section*{Acknowledgements}

Thanks to the DoE and National Georgraphic for images.
Thanks James Adams and James Geddes at the Turing Institute for several helpful pointers to 
background on bio- \& crypto- technologies.

\bibliographystyle{ACM-Reference-Format} 
\bibliography{hotbio}

\end{document}